\documentclass[pra,groupedaddress,nobalancelastpage,twocolumn]{revtex4}
\usepackage{graphicx}
\usepackage{amsmath}
\usepackage{amsfonts}
\usepackage{amssymb}

\begin{document}

\title{Light reflection upon a movable mirror as a paradigm for continuous variable
teleportation network}
\author{Stefano Pirandola, Stefano Mancini, David Vitali, and Paolo Tombesi}

\affiliation{
INFM, Dipartimento di Fisica,
Universit\`a di Camerino, I-62032 Camerino, Italy}

\begin{abstract}
We present an optomechanical system as a paradigm of three-mode teleportation
network. Quantum state transfer among optical and vibrational modes becomes
possible by exploiting correlations established by radiation pressure.
\end{abstract}
\maketitle

\affiliation{
INFM, Dipartimento di Fisica,
Universit\`a di Camerino, I-62032 Camerino, Italy}

\section{Introduction}

Quantum teleportation is one of the most fascinating possibilities offered by
quantum information processing \cite{man}. In the standard protocol, Alice is
to transfer an unknown quantum state to Bob using, as the sole resources, some
previous shared entanglement (quantum channel) and a classical channel capable
of communicating measurement results. Often the quantum channel for continuous
variable (CV) teleportation is realized by using two-mode squeezed (TMS)
states which may mimic nonclassical Einstein-Podolsky-Rosen (EPR) correlations
\cite{vai}. The concept of quantum channel can be generalized to more than two
parties to form a teleportation network. In such a case it can be used for
telecloning \cite{Tclon2}, or, again, for teleportation by distilling
bipartite entanglement \cite{telenet}.

As a consequence of a famous science fiction saga, a wish to involve
macroscopic systems into teleportation protocols now hovers in people mind.
Recently, we have gone along this way showing the possibility to teleport a
quantum state onto the vibrational mode of a movable mirror \cite{PRA}. In
doing so we have exploited an intense radiation field impinging on the movable
mirror and reflected into an intense elastic carrier together with anelastic
sideband modes. The latter together with the mirror vibrational mode
constitute a tripartite quantum system. Here we shall examine it as a
teleportation network where we exploit the bipartite entanglement, and more
strongly the EPR correlations, which can be distilled for certain groupings of
two modes, simply tracing out the remaining mode or measuring it by heterodyne
detection and communicating the result. In both situations the distilled
channel comes out as a (nonzero mean) Gaussian bipartite state. The model is
presented in Section \ref{Hami}. Then, we report the general teleportation
protocol through a Gaussian channel in Section \ref{teoria}. Hence, we examine
the two situations above in Sections \ref{appli1} and \ref{appli2}. Section
\ref{conclu} is for conclusions.

\section{Effective Hamiltonian and system dynamics}

\label{Hami}

\begin{figure}[th]
\begin{center}
\includegraphics[width=2.2in]{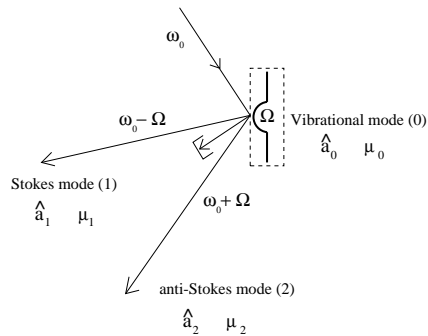}
\end{center}
\caption{{\small Scheme of the optomechanical system.}} \label{schema1}
\end{figure}

We consider a perfectly reflecting mirror and an intense quasi-monochromatic
laser beam impinging on its surface (see Fig.~\ref{schema1}). The laser beam
is linearly polarized along the mirror surface and focused in such a way as to
excite Gaussian acoustic modes of the mirror. These modes describe small
elastic deformations of the mirror along the direction orthogonal to its
surface and are characterized by a small waist, a large quality factor and a
small effective mass \cite{PIN99}. It is possible to adopt a single
vibrational mode description limiting detection bandwidth to include a single
mechanical resonance of frequency $\Omega$. In this description the incident
laser beam, with frequency $\omega_{0}$, is reflected into an elastic carrier
mode, with the same frequency $\omega_{0}$, and two additional weak anelastic
sideband modes with frequencies $\omega_{0}\pm\Omega$. The physical process is
very similar to a stimulated Brillouin scattering, even though in this case
the Stokes and anti-Stokes component are back-scattered by the acoustic wave
at reflection, and the optomechanical coupling is provided by the radiation
pressure. Treating classically the intense incident beam (and the carrier
mode), the quantum system is composed by three interacting quantized bosonic
modes, i.e. the vibrational mode and the two sideband modes. In our
description, vibrational, Stokes and anti-Stokes modes are denoted as $0,1,$
and $2$ respectively. In general the $k^{th}$ mode ($k=0,1,2$)\ is
characterized by ladder operators $\hat{a}_{k},\hat{a}_{k}^{\dagger}$ and by
quadrature operators $\hat{X}_{k}=(\hat{a}_{k}+\hat{a}_{k}^{\dagger})/\sqrt
{2}$, $\hat{P}_{k}=(\hat{a}_{k}-\hat{a}_{k}^{\dagger})/i\sqrt{2}$ ($[\hat
{X}_{k},\hat{P}_{k}]=i$). In \cite{PRA,JOPB}\ we have derived an effective
interaction Hamiltonian for that system
\begin{equation}
{\hat{H}}_{eff}=-i\hbar\chi({\hat{a}}_{1}{\hat{a}}_{0}-{\hat{a}}_{1}^{\dag
}{\hat{a}}_{0}^{\dag})-i\hbar\theta({\hat{a}}_{2}{\hat{a}}_{0}^{\dag}-{\hat
{a}}_{2}^{\dag}{\hat{a}}_{0})\,, \label{eq:Heff}
\end{equation}
where $\chi$ and $\theta$ are couplings constants whose ratio
$r\equiv \theta/\chi=\left[
(\omega_{0}+\Omega)/(\omega_{0}-\Omega)\right]  ^{1/2} \geq1$ only
depends on the involved frequencies. The system dynamics is
satisfactorily reproduced by the Hamiltonian of
Eq.~(\ref{eq:Heff}) as long as the dissipative coupling of the
mirror vibrational mode with its environment is negligible. This
happens if the interaction time is much smaller than the
relaxation time of the vibrational mode (which can be $\sim1$ s
\cite{TIT99}) and therefore means having a high-Q vibrational mode
(typically $\Omega\sim$ MHz).

The dynamics can be easily studied in terms of the symmetrically
ordered characteristic function $\Phi(\mu),$
$\mu\equiv(\mu_{0},\mu_{1},\mu_{2})$ \cite{QO94}, where
$\mu_{k}=\mu_{k}^{(R)}+i\mu_{k}^{(I)}$ is a complex variable
corresponding to the operator ${\hat{a}}_{k}$ ($k=0,1,2$). The
relation between the density operator and the corresponding
characteristic function is given by
$\rho_{012}=\pi^{-3}[\prod_{k=0}^{2}\int d^{2}\mu_{k}
\hat{D}_{k}^{\dagger}(\mu_{k})]\Phi(\mu)$, where
$\hat{D}_{k}(\mu_{k})=\exp\{\mu_{k}\hat{a}_{k}^{\dagger}-\mu_{k}^{\ast}\hat
{a}_{k}\}=\exp\{i\sqrt{2}(\mu_{k}^{(I)}\hat{X}_{k}-\mu_{k}^{(R)}\hat{P}
_{k})\}$. Notice that the Fourier transform defined as above
through the displacement operators $\hat{D}_{k}(\mu_{k})$, creates the correspondence
$\mu_{k}^{(I)} \leftrightarrow\hat{X}_{k}$,
$-\mu_{k}^{(R)}\leftrightarrow\hat{P}_{k}$. From
Eq.~(\ref{eq:Heff}) we can deduce the dynamical equation for
$\Phi(\mu,t)$. If we assume the initial condition
$\Phi(\mu,t=0)=\exp\left[
-\overline{n}|\mu_{0}|^{2}-\sum_{k=0}^{2}|\mu_{k}|^{2}/2\right]
$, corresponding to a vacuum state for modes $1,2$ and to a
thermal state with mean thermal number of excitations $\bar{n}$
for mode $0$, the total state at time $t$ is Gaussian, with
characteristic function
\begin{align}
&
\Phi(\mu,t)=\exp[-Q_{0}|\mu_{0}|^{2}-Q_{1}|\mu_{1}|^{2}-Q_{2}|\mu_{2}
|^{2}+T_{0}(\mu_{1}\mu_{2}\nonumber\\
&
+\mu_{1}^{\ast}\mu_{2}^{\ast})+T_{1}(\mu_{0}\mu_{2}^{\ast}+\mu_{0}^{\ast
}\mu_{2})+T_{2}(\mu_{1}\mu_{0}+\mu_{1}^{\ast}\mu_{0}^{\ast})],
\label{eq:newtotstate}
\end{align}
where: $Q_{0}\equiv\mathcal{B}+1/2$, $Q_{1}\equiv\mathcal{A}+1/2$,
$Q_{2}\equiv\mathcal{E}+1/2$, $T_{0}\equiv\mathcal{F}$, $T_{1}\equiv
\mathcal{D}$, $T_{2}\equiv\mathcal{C}$ and the coefficients $\mathcal{A}$,
$\mathcal{B}$, $\mathcal{C}$, $\mathcal{D}$, $\mathcal{E}$ and $\mathcal{F}$,
explicitly given in \cite{JOPB}, depend upon $\bar{n}$, $r$ and the scaled
time $t^{\prime}\equiv t\sqrt{\theta^{2}-\chi^{2}}$. Since the state of
Eq.~(\ref{eq:newtotstate}) is a zero-mean Gaussian state, it can be fully
described by its correlation matrix (CM) $V$, defined by $V_{lm}\equiv
\langle\Delta\hat{\xi}_{l}\Delta\hat{\xi}_{m}+\Delta\hat{\xi}_{m}\Delta
\hat{\xi}_{l}\rangle/2$ ($l,m=1,...,6$),\ where $\Delta\hat{\xi}_{l}\equiv
\hat{\xi}_{l}-\langle\hat{\xi}_{l}\rangle$ and $\hat{\xi}$ denotes the vector
of quadratures: $\hat{\xi}\equiv(\hat{X}_{0},\hat{P}_{0},\hat{X}_{1},\hat
{P}_{1},\hat{X}_{2},\hat{P}_{2})$. In fact, introducing the real vectors
$\vec{\mu}_{k}$ ($k=0,1,2$), defined by
\begin{equation}
\mu_{k}=\mu_{k}^{(R)}+i\mu_{k}^{(I)}\longleftrightarrow(\mu_{k}^{(I)},-\mu
_{k}^{(R)})\equiv\vec{\mu}_{k}\in\mathbb{R}^{2},
\label{eq:realvar1}
\end{equation}
so that $\mathbb{C}^{3}\ni\mu\equiv(\mu_{0},\mu_{1},\mu_{2}
)\longleftrightarrow(\vec{\mu}_{0},\vec{\mu}_{1},\vec{\mu}_{2})\equiv\vec{\mu
}\in\mathbb{R}^{6}$, and expressing $\Phi(\mu,t)$ in terms of
$\vec{\mu}$, we get from Eq.~(\ref{eq:newtotstate})
\begin{equation}
\Phi(\vec{\mu},t)=e^{-\vec{\mu}V\vec{\mu}^{T}}\,, \label{symme}
\end{equation}
where the CM $V$ appears and it is explicitly given by
\begin{equation}
V=\left(
\begin{tabular}
[c]{cc|cc|cc}
$Q_{0}$ & $0$ & $T_{2}$ & $0$ & $-T_{1}$ & $0$\\
$0$ & $Q_{0}$ & $0$ & $-T_{2}$ & $0$ & $-T_{1}$\\\hline
$T_{2}$ & $0$ & $Q_{1}$ & $0$ & $T_{0}$ & $0$\\
$0$ & $-T_{2}$ & $0$ & $Q_{1}$ & $0$ & $-T_{0}$\\\hline
$-T_{1}$ & $0$ & $T_{0}$ & $0$ & $Q_{2}$ & $0$\\
$0$ & $-T_{1}$ & $0$ & $-T_{0}$ & $0$ & $Q_{2}$
\end{tabular}
\right)  \;. \label{CMgrande}
\end{equation}
Since we deal with a closed system of three interacting
oscillators its dynamics is periodic in $t'$ with period $2\pi$. The separability properties
of the tripartite system $\rho _{012}$ and those of the bipartite
reduced systems $\rho^{(k)}=tr_{k} (\rho_{012})$ ($k=0,1,2)$, have
been already studied in \cite{PRA}. Here we briefly recall these
properties referring to \cite{entcirac}\ for the definition of
entanglement classes of a tripartite CV Gaussian state. The state
$\rho_{012}$ is almost everywhere fully entangled (\textit{class
1}) except for isolated times $t^{\prime}=2m\pi$
($m\in\mathbb{N}$) when it is fully separable (\textit{class 5})
and $t^{\prime}=(2m+1)\pi$ when it is one-mode biseparable
(\textit{class 2}). In the latter case, it is equal to the tensor
product of a TMS state for the optical modes and of a thermal
state for the vibrational mode.

The inseparability between one of the three modes and the other two parties
can be exploited to implement a telecloning protocol \cite{Tclon2}. When we
trace out one of the three modes, we can distill bipartite entanglement only
between modes 1 and 2 or between modes 1 and 0. In the first case (1 and 2),
the entanglement between the optical modes exists at all times (except
$t^{\prime}=2m\pi$), and it is extremely robust with respect to temperature.
Such modes show also robust EPR correlations which are temperature-independent
at $t^{\prime}=(2m+1)\pi$ when the state is a TMS state \cite{JOPB}. In the
second case, the entanglement between the optical Stokes mode and the mirror
vibrational mode exists in two limited time intervals, just after $t^{\prime
}=0$ and just before $t^{\prime}=2\pi$, and they become narrower by increasing
the temperature. In \cite{PRA} the second time interval has been exploited to
realize quantum teleportation from an optical to the mirror vibrational mode.

\section{\label{teoria}CV teleportation through an EPR channel}

Suppose to have a quantum teleportation network, i.e. a quantum channel given
by a truly multipartite entangled state shared among $N$ parties. Such a
situation has been studied in \cite{telenet} where, from a particular
$N-$partite entangled state, a bipartite entanglement between any two of the
$N$ parties can be distilled to enable quantum teleportation. Clearly the
distillability of bipartite entanglement from the total channel is a necessary
condition to make the teleportation network really \textit{quantum} (i.e. not
reproducible by any local classical means).

\begin{figure}[th]
\begin{center}
\includegraphics[width=2.2in]{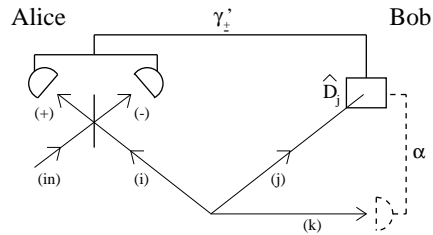}
\end{center}
\caption{{\small Three-mode teleportation network. In one case
mode $k$ is simply traced out while in the other case it is
heterodyned and the measurement result $\alpha$ sent to Bob. From
a zero-mean Gaussian state for $i,j,k$ we always distill a
bipartite Gaussian state. In the first case the quantum channel
has zero drift and Bob performs the displacement $\alpha
_{j}\longrightarrow\alpha_{j}-\gamma_{\pm}^{\prime}$. In the
second case, the channel has a nonzero drift, known from the knowledge of $\alpha$, and
Bob performs a modified displacement
$\alpha_{j}\longrightarrow\alpha_{j}-\gamma_{\pm}^{\prime}
-\delta_{\pm}$ with $\delta_{\pm}$ depending on the drift (see
text).}} \label{schema2}
\end{figure}

Here we consider a distilled channel consisting of a bipartite
Gaussian state with a \textit{known} drift. That happens for
example when, from a $N$-partite zero-mean Gaussian state, we
trace out $N-2$ modes, or when we measure them by heterodyne
detection (communicating the results to Bob through a classical
channel). Suppose that such a distilled channel is shared in a
network by Alice (mode $i$) and Bob (mode $j$) (for simplicity
consider $N=3$ and see Fig.~\ref{schema2}). If their channel is
separable then they cannot perform quantum teleportation but, if
it is entangled, then they can try to perform it by exploiting
possible EPR correlations of their channel. We define
``EPR$+$'' or ``EPR$-$'' correlations in the following way
\begin{equation}
\text{EPR}\pm \Leftrightarrow \langle\Delta
\hat{X}_{\pm}^{2}\rangle +\langle\Delta
\hat{P}_{\mp}^{2}\rangle<2, \label{eq:EPRcond2}
\end{equation}
where $\hat{X}_{\pm}\equiv(\hat{X}_{i}\pm\hat{X}_{j})$ and $\hat{P}_{\pm
}\equiv(\hat{P}_{i}\pm\hat{P}_{j})$. Depending on the supposed EPR
correlations, Alice and Bob can implement an appropriate teleportation
protocol. We treat both cases in a compact form adopting the phase-space
approach of \cite{CHI02}. The distilled channel is described by a Gaussian
Wigner function $W^{ch}(\alpha_{i},\alpha_{j})$ with a known drift and
supposing to possess EPR$\pm$ correlations according to definition
(\ref{eq:EPRcond2}). The unknown input state at Alice's station is described
by $W^{in}(\gamma)$ so that the total state before the beam-splitter is given
by $W(\gamma,\alpha_{i},\alpha_{j})=W^{in}(\gamma)W^{ch}(\alpha_{i},\alpha
_{j})$. After the beam splitter (Fig.~\ref{schema2})\ we have
\begin{equation}
W(\nu_{+},\nu_{-},\alpha_{j})=W^{in}\left(
\frac{\nu_{+}-\nu_{-}}{\sqrt{2} }\right)  W^{ch}\left(
\frac{\nu_{+}+\nu_{-}}{\sqrt{2}},\alpha_{j}\right),
\label{eq:Wtotstate}
\end{equation}
where $\nu_{\pm}=(\alpha_{i}\pm\gamma)/\sqrt{2}$ are the complex
amplitudes of the output modes $\pm$ having quadratures
$\hat{x}_{\pm}=(\hat{X}_{i}\pm \hat{X}_{in})/\sqrt{2}$ and
$\hat{p}_{\pm}=(\hat{P}_{i}\pm\hat{P}_{in} )/\sqrt{2}$. Depending
on the (supposed) EPR correlations of the channel, Alice measures
an appropriate pair of quadratures of the output modes, namely
EPR$\pm$ implies measuring $(\hat{x}_{\pm},\hat{p}_{\mp})$. In
terms of complex amplitudes, Alice's measurement results
$(x_{\pm},p_{\mp})$ are expressed by the variable
$\gamma_{\pm}^{\prime}$ with probability density
$P(\gamma_{\pm}^{\prime})$ where
\[
\gamma_{+}^{\prime}\equiv\sqrt{2}(\nu_{+}^{(R)}-i\nu_{-}^{(I)})\text{ \ and
\ }\gamma_{-}^{\prime}\equiv\sqrt{2}(i\nu_{+}^{(I)}-\nu_{-}^{(R)}).
\]
Alice's detection collapses the total state (\ref{eq:Wtotstate}) whence Bob's
conditioned state results
\begin{equation}
W(\alpha_{j}|\gamma_{\pm}^{\prime})=\frac{1}{P(\gamma_{\pm}^{\prime})}\int
d^{2}\gamma
W^{in}(\gamma)W^{ch}[\pm(\gamma_{\pm}^{\prime\ast}-\gamma^{\ast
}),\alpha_{j}].\label{eq:WBobcon}
\end{equation}
State (\ref{eq:WBobcon}) is exactly the state at Bob's station when Bob
receives from Alice her result $\gamma_{\pm}^{\prime}$ through a classical
channel. At this point Bob completes the teleportation protocol by a suitable
displacement of his mode, namely
\begin{equation}
\alpha_{j}\rightarrow\alpha_{j}-\gamma_{\pm}^{\prime}-\delta_{\pm
},\label{eq:Bobdisp}
\end{equation}
where the first quantity $\gamma_{\pm}^{\prime}$ balances Alice's
detection while the second one $\delta_{\pm}$ (to be specified)
balances the drift of the channel and depends on the type of EPR
correlations, too. The final state at Bob's station is then given
by $W(\alpha_{j}-\gamma_{\pm}^{\prime}
-\delta_{\pm}|\gamma_{\pm}^{\prime})$. In order to get the
teleportation fidelity, we compute the mean state teleported to
Bob by averaging over all possible results $\gamma_{\pm}^{\prime}$
\begin{align}
W_{\pm}^{out}(\alpha_{j};\delta_{\pm}) &  =\int d^{2}\gamma_{\pm}^{\prime
}P(\gamma_{\pm}^{\prime})W(\alpha_{j}-\gamma_{\pm}^{\prime}-\delta_{\pm
}|\gamma_{\pm}^{\prime})\nonumber\\
&  =\int d^{2}\gamma K_{\pm}(\alpha_{j},\gamma;\delta_{\pm})W^{in}
(\gamma),\label{eq:Wout2}
\end{align}
where
\begin{equation}
K_{\pm}(\alpha_{j},\gamma;\delta_{\pm})\equiv\int
d^{2}\gamma_{\pm}^{\prime
}W^{ch}[\pm(\gamma_{\pm}^{\prime\ast}-\gamma^{\ast}),\alpha_{j}-\gamma_{\pm
}^{\prime}-\delta_{\pm}] \label{eq:Kernel}
\end{equation}
depends also on the parameter $\delta_{\pm}$.
The input-output relation (\ref{eq:Wout2}) can be greatly simplified if we
introduce the symmetrically ordered characteristic functions given by the
Fourier transform of the Wigner functions, i.e. $\Phi(\lambda)=\int
d^{2}\alpha e^{\lambda\alpha^{\ast}-\lambda^{\ast}\alpha}W(\alpha)$ where
$\lambda=\lambda^{(R)}+i\lambda^{(I)}$ is the conjugate variable of the
complex amplitude $\alpha$. In fact, applying the Fourier transform to
(\ref{eq:Wout2}), we get:
\begin{equation}
\Phi_{\pm}^{out}(\lambda;\delta_{\pm})=e^{\lambda\delta_{\pm}^{\ast}
-\lambda^{\ast}\delta_{\pm}}\Phi^{ch}(\mp\lambda^{\ast},\lambda)\Phi
^{in}(\lambda).\label{eq:inout1}
\end{equation}
where $\Phi^{in} (\mu_{in}),\Phi^{out}(\mu_{j})$ and $\Phi^{ch}(\mu_{i},\mu_{j})$
are evaluated according to the substitutions shown in Eq.~(\ref{eq:inout1}).
If we consider a pure
input state then the fidelity takes the form
\begin{equation}
F_{\pm}(\delta_{\pm})=\frac{1}{\pi}\int d^{2}\lambda\left|
\Phi^{in} (\lambda)\right|
^{2}[\Phi^{ch}(\mp\lambda^{\ast},\lambda)]^{\ast}
e^{-\lambda\delta_{\pm}^{\ast}+\lambda^{\ast}\delta_{\pm}},\label{eq:espfid2}
\end{equation}
where we have used Eq.~(\ref{eq:inout1}). In particular we
consider a Gaussian pure input state and it can always be chosen
with zero-mean, because of the invariance of
$F_{\pm}(\delta_{\pm})$ with respect to the amplitude of the input
state. Using the real variables of Eq.~(\ref{eq:realvar1}), the
characteristic functions of the input and the channel are given
respectively by
$\Phi^{in}(\vec{\mu}_{in})=\exp(-\vec{\mu}_{in}V^{in}\vec{\mu}_{in}^{T})$
and
$\Phi^{ch}(\vec{\mu}_{i},\vec{\mu}_{j})=\exp[-(\vec{\mu}_{i},\vec{\mu}
_{j})V^{ch}(\vec{\mu}_{i},\vec{\mu}_{j})^{T}+D^{ch}(\vec{\mu}_{i},\vec{\mu
}_{j})^{T}]$, where the CMs $V^{in}$ and $V^{ch}$ are
real positive matrices, while the drift vector $D^{ch}$ has
pure imaginary elements. In particular we can write:
\begin{align}
V^{ch} &  \equiv\left(
\begin{array}
[c]{cc}
A & C\\
C^{T} & B
\end{array}
\right)  \nonumber\\
D^{ch} &  \equiv2i\left(  d_{1}\,,\,d_{2}\,,\,d_{3}\,,\,d_{4}\right)
\nonumber
\end{align}
with $d_{k}\in\mathbb{R}$. Inserting $\Phi^{in}(\vec{\mu}_{in})$ and
$\Phi^{ch}(\vec{\mu}_{i},\vec{\mu}_{j})$ in Eq.~(\ref{eq:espfid2}), we get:
\begin{equation}
F_{\pm}(\delta_{\pm})=\frac{1}{\sqrt{\det(E_{\pm})}}\exp(-Q_{\pm
})\label{eq:Fidgen}
\end{equation}
where
\begin{align}
E_{\pm} &  \equiv2V^{in}+RAR+B\pm RC\pm C^{T}R^{T}\label{eq:termE}\\
R &  \equiv\left(
\begin{array}
[c]{cc}
1 & 0\\
0 & -1
\end{array}
\right)  \label{eq:MatrixR}
\end{align}
and
\begin{align}
Q_{\pm} &  \equiv D_{\pm}(E_{\pm})^{-1}(D_{\pm})^{T}\geq0\label{eq:termQ}\\
D_{\pm} &  \equiv(-\delta_{\pm}^{(R)}\mp
d_{1}-d_{3},-\delta_{\pm}^{(I)}\pm d_{2}-d_{4}).\label{eq:termD}
\end{align}
In Eq.~(\ref{eq:Fidgen}) the positive matrix $E_{\pm}$ depends
only\ upon the CMs $V^{in},V^{ch}$ through (\ref{eq:termE}), while
the non-negative term $Q_{\pm}$ is linked also to the vector
$D_{\pm}$ through (\ref{eq:termQ}), which, in turn, contains both
the drift of the Gaussian channel and the drift created by Bob's
additional displacement $\delta_{\pm}=\delta_{\pm}
^{(R)}+i\delta_{\pm}^{(I)}$. Without such a displacement (i.e.
$\delta_{\pm }=0$), the teleported state acquires a nonzero drift
from the channel and the corresponding fidelity $F_{\pm}(0)$ will
depend upon such a drift via a decreasing exponential. This does
not happen if $D^{ch}=0$, or, more generally, if
$d_{1}-d_{3}=d_{2}-d_{4}=0$. In the general case, the only way in
which Bob can eliminate the nonzero drift is to choose an
additional displacement $\delta_{\pm}$ which perfectly cancels the
effects of $D^{ch}$, i.e., is such that $D_{\pm}=0$. Using
(\ref{eq:termD}), that means:
\begin{equation}
\delta_{\pm}^{(R)}=\mp d_{1}-d_{3}\text{ \ and \
}\delta_{\pm}^{(I)}=\pm d_{2}-d_{4}.\label{eq:termdelt2}
\end{equation}
With such a displacement the teleportation becomes independent from the channel
drift and the fidelity (\ref{eq:Fidgen}) becomes
\begin{equation}
F_{\pm}=\frac{1}{\sqrt{\det(E_{\pm})}}.\label{eq:Fidgen2}
\end{equation}
Suppose now that our input state is a coherent state and that our quantum
channel has the CM in the standard form
\begin{equation}
V^{ch}=\left(
\begin{array}
[c]{cccc}
a & 0 & c & 0\\
0 & a & 0 & c^{\prime}\\
c & 0 & b & 0\\
0 & c^{\prime} & 0 & b
\end{array}
\right) ; \label{eq:stform}
\end{equation}
then the fidelity (\ref{eq:Fidgen2}) is simply related to the quadrature
variances:
\begin{equation}
F_{\pm}=\left[  \left(
1+\langle\Delta\hat{X}_{\pm}^{2}\rangle\right) \left(
1+\langle\Delta\hat{P}_{\mp}^{2}\rangle\right)  \right]
^{-1/2},\label{eq:FIDstform}
\end{equation}
where explicitly $\langle\Delta\hat{X}_{\pm}^{2}\rangle=a+b\pm2c$ and
$\langle\Delta\hat{P}_{\pm}^{2}\rangle=a+b\pm2c^{\prime}$. In particular, if
$c^{\prime}=c$, then $\langle\Delta\hat{X}_{\pm}^{2}\rangle=\langle\Delta
\hat{P}_{\pm}^{2}\rangle\geq1$ and $F_{\pm}\leq1/2=F_{class}$, i.e., quantum
teleportation is not possible. Instead if $c^{\prime}=-c$ then we have
$\langle\Delta\hat{X}_{\pm}^{2}\rangle=\langle\Delta\hat{P}_{\mp}^{2}\rangle$
and $F_{\pm}=\left[  1+\langle\Delta\hat{X}_{\pm}^{2}\rangle\right]  ^{-1}$.
In such a case the existence of EPR correlations (\ref{eq:EPRcond2}) is
equivalent to quantum teleportation, i.e.
\begin{equation}
\langle\Delta\hat{X}_{\pm}^{2}\rangle=\langle\Delta\hat{P}_{\mp}^{2}
\rangle<1\Leftrightarrow
F_{\pm}>\frac{1}{2}=F_{class}.\label{eq:EPReqTEL}
\end{equation}
In this sense the Gaussian state
$\Phi^{ch}(\vec{\mu}_{i},\vec{\mu} _{j})\longleftrightarrow
V^{ch},D^{ch}$\ having the CM of Eq.~(\ref{eq:stform} ) with
$c^{\prime}=-c$ is an \textit{EPR channel} generalizing the
two-parameter EPR channel studied in \cite{BRA99}. Note that the
left hand side of Eq.~(\ref{eq:EPReqTEL}) is a sufficient
condition for bipartite entanglement \cite{PRLdu}.

\section{\label{appli1}Tracing out one mode}

\begin{figure}[th]
\begin{center}
\includegraphics[width=2.7in]{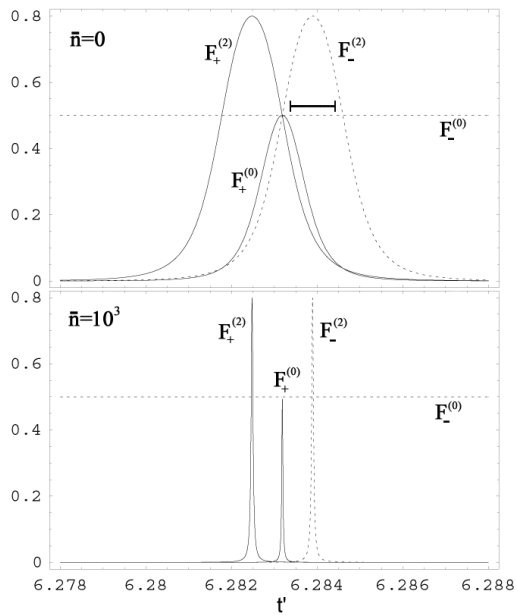}
\end{center}
\caption{{\small Traced out mode case. Fidelities
$F_{\pm}^{(0)},F_{\pm }^{(2)}$\ versus $t^{\prime}$ (around
$t^{\prime}=2\pi$) for $\bar{n}=0$ and $\bar{n}=10^{3}$. We have
set $r=1+2.5\times10^{-7}$. The telecloning time interval has been
marked.}} \label{fig0}
\end{figure}

We can consider our optomechanical system (Fig.~\ref{schema1}) as a 3-mode
teleportation network where we distill a quantum channel between two arbitrary
modes $i$ and $j$ tracing out the remaining mode $k$ (Fig.~\ref{schema2}). We
denote with $\rho^{(k)}=tr_{k}(\rho_{012})$ the reduced state. The
corresponding characteristic function is given by $\Phi^{(k)}(\mu_{i},\mu
_{j})=\Phi(\mu_{i},\mu_{j},\mu_{k}=0)$, and using Eq.~(\ref{eq:realvar1}), we
obtain $\Phi^{(k)}(\vec{\mu}_{i},\vec{\mu}_{j})=\exp[-(\vec{\mu}_{i},\vec{\mu
}_{j})V^{(k)}(\vec{\mu}_{i},\vec{\mu}_{j})^{T}]$, where:
\begin{equation}
V^{(k)}=\left(
\begin{array}
[c]{cccc}
Q_{i} & 0 & (-)^{k}T_{k} & 0\\
0 & Q_{i} & 0 & -T_{k}\\
(-)^{k}T_{k} & 0 & Q_{j} & 0\\
0 & -T_{k} & 0 & Q_{j}
\end{array}
\right)  \label{eq:CMVk}
\end{equation}
In other words the distilled channel is a zero-mean Gaussian state
with CM in the standard form (\ref{eq:CMVk}). It follows that
Alice ($i$) and Bob ($j$) can arrange the teleportation protocol
of Section \ref{teoria} without any additional displacement
$\delta_{\pm}$ and with fidelity $F_{\pm}^{(k)}$ given by
Eq.~(\ref{eq:FIDstform}) for a coherent input. It is easy to see
from Eq.~(\ref{eq:CMVk})\ that for $k=1$ we have
$F_{\pm}^{(1)}\leq1/2$ while for $k=0,2$ we have an EPR channel
with fidelity $F_{\pm}^{(k)}=[1+Q_{i}+Q_{j} \pm2T_{k}]^{-1}$. In
Fig.~\ref{fig0} we report $F_{\pm}^{(0)},F_{\pm}^{(2)}$ versus
$t^{\prime}$ for $\bar{n}=0,10^{3}$ and choosing
$r=1+2.5\times10^{-7}$ as in Ref.~\cite{PRA}. We have
$F_{\pm}^{(0)}(2\pi)=1/2$ but while $F_{+}^{(0)}\leq1/2$ $\forall
t^{\prime}$, we have $F_{-}^{(0)}\geq1/2$ $\forall t^{\prime}$ (at
least up to $\bar{n}=10^{5}$, see also Fig.~\ref{fig2}). In
particular, $F_{-}^{(0)}$ is approximately constant near $2\pi$
and has the expansion $F_{-}^{(0)}\sim\lbrack2-(t^{\prime}-2\pi
)^{2}/2]^{-1}$. Analyzing $F_{\pm}^{(2)}$, we see that the quantum
channel distilled for the Stokes mode (1) and the mirror mode (0)
has EPR$+$ correlations before $2\pi$ (exploited in \cite{PRA})\
and EPR$-$ correlations after $2\pi$. The existence of this
symmetric region is important not only because it allows to
perform an additional quantum teleportation just after
$t^{\prime}=2\pi$ ($\Leftrightarrow$ $t^{\prime}=0$) but also
because in the time
interval delimited by the intersections between $F_{-}^{(2)}$ and $F_{-}^{(0)}$ (see Fig.~\ref{fig2})
these latter fidelities are strictly greater than $1/2$. This means that a region exists after $2\pi$ where
a telecloning protocol can be arranged using the Stokes mode (1)
as ''port'' and the anti-Stokes (2) and the mirror (0) modes as
''receivers'' [Alice (1) measures $(\hat{x}_{-},\hat{p}_{+})$ and
classically\ sends the result $\gamma_{-}^{\prime}$ to Bob (0) and
Charlie (2), who perform the displacements
$\alpha_{j}\rightarrow\alpha_{j} -\gamma_{-}^{\prime}$,
$j=0,2$]. The clones created at the receivers are both quantum,
but the asymmetry of the total channel is such that $F_{-}^{(0)}$
is only slightly greater than $1/2$ in all the ``telecloning
interval'', while $F_{-}^{(2)}$ takes its maximum
$[F_{-}^{(2)}]_{\max}=(1+r)^{2} /[1+2r(1+r)]\simeq0.8$,
$\forall\bar{n}$, and for $t_{\max}^{\prime} =\varsigma/2+2\pi$
with $\varsigma\equiv\cos^{-1}(2r^{-2}-1)$. Since
$F_{-}^{(2)}(t_{\max}^{\prime}-\varsigma/2)=F_{-}^{(2)}(t_{\max}^{\prime
}+\varsigma/2)=(2+\bar{n})^{-1}$, the telecloning interval becomes
narrower by increasing the temperature. On the other hand, the
behavior of $F_{-}^{(0)}$ (see $F_{-}^{(0,tr)}$ in
Fig.~\ref{fig2}) shows the robustness with respect to temperature
of the EPR$-$ correlations relative to the distilled channel
between the two optical modes: the curves for $\bar{n}=0$ and
$\bar{n}=10^{5}$ are almost indistinguishable. Such EPR$-$
channel, which reduces to a TMS state at $t^{\prime}=\pi$
\cite{JOPB} (it is in particular $F_{-}^{(0)}
(\pi)=1/2+r/(r^{2}+1)\simeq1$), enables quantum teleportation even
at high temperatures and points out the present optomechanical
system as an alternative source of two-mode squeezing. From the
$r$-dependence of the maximum values $[F_{-}^{(2)}]_{\max}$ and
$F_{-}^{(0)}(\pi)$, we can see that the optimal values, achieved
for $r\rightarrow1$, are $4/5$ and $1$ respectively.

\section{\label{appli2}Heterodyning one mode}

\begin{figure}[th]
\begin{center}
\includegraphics[width=3in]{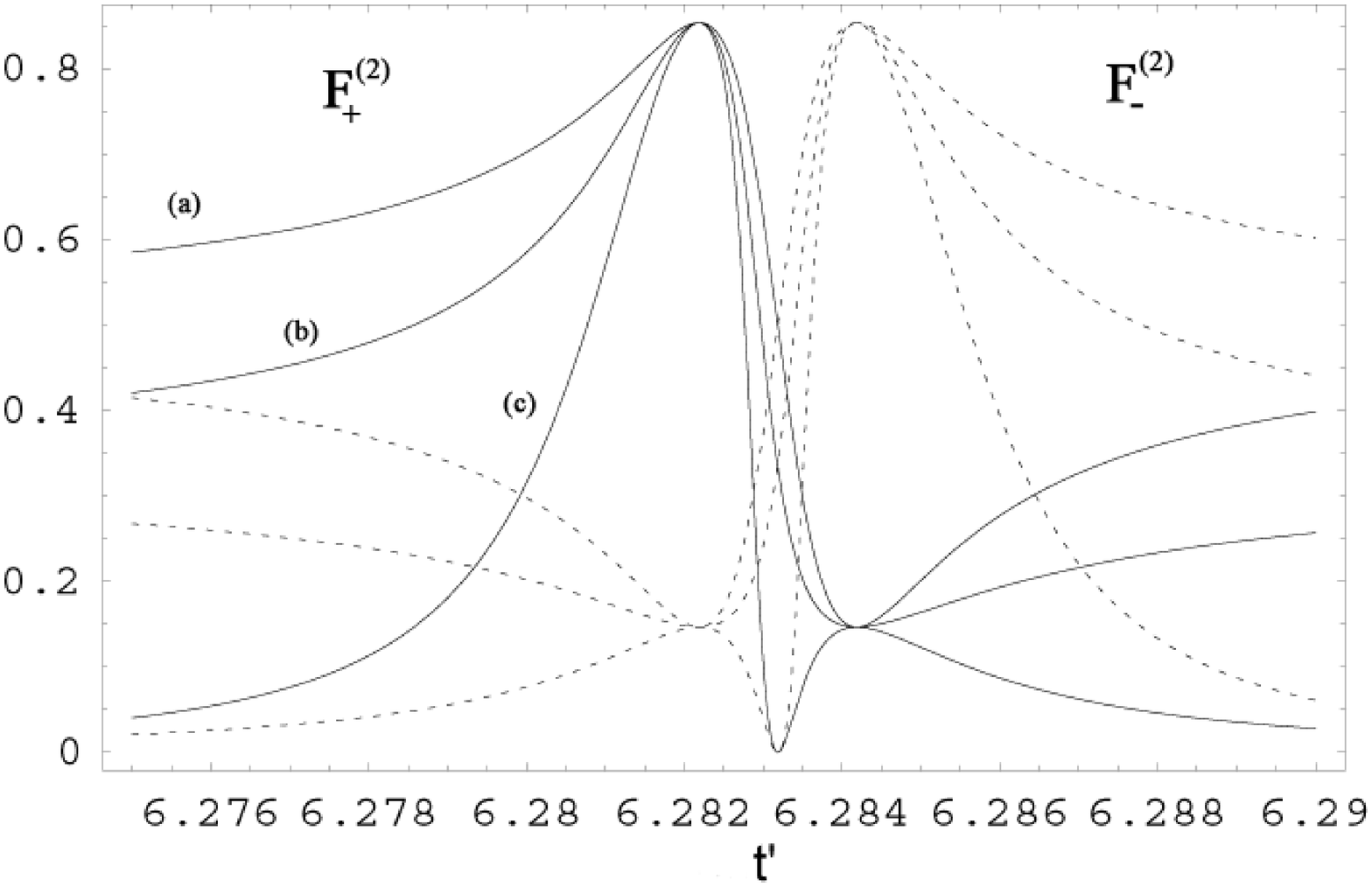}
\end{center}
\caption{{\small Heterodyne detection case. Fidelities
$F_{\pm}^{(2)} $\ versus $t^{\prime}$ (around $t^{\prime}=2\pi$)
for $\bar{n}=0$ (a), $\bar{n}=1$ (b) and $\bar{n}=10^{7}$ (c). We
have set $r=1+2.5\times10^{-7}$ .}} \label{fig1}
\end{figure}

\begin{figure}[th]
\begin{center}
\includegraphics[width=2.7in]{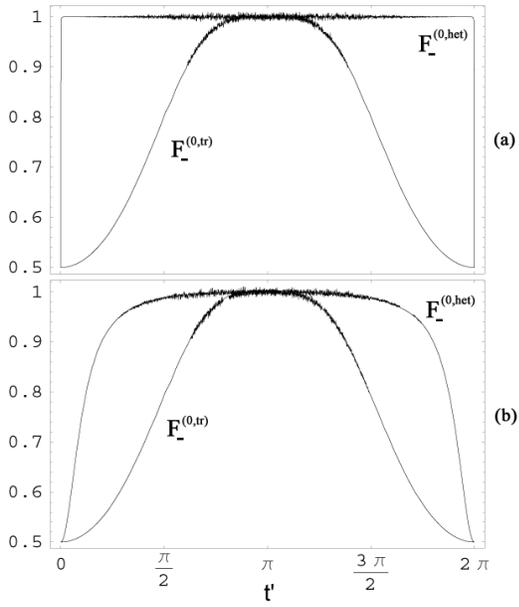}
\end{center}
\caption{{\small Fidelity $F_{-}^{(0)}$ versus $t^{\prime}$ in the
traced out mode case $F_{-}^{(0,tr)}$ and in the detected mode
case $F_{-}^{(0,het)} $, for $\bar{n}=0$ (a) and $\bar{n}=10^{5}$
(b). We have set $r=1+2.5\times 10^{-7}$.}} \label{fig2}
\end{figure}

Consider now our optomechanical system as a 3-mode teleportation
network where we distill a quantum channel between two arbitrary
modes $i$ and $j$ heterodyning the remaining mode $k$ and sending
the result to Bob through a classical channel
(Fig.~\ref{schema2}). We denote with $\alpha$ the measurement
result and again with $\rho^{(k)}$ the state of the reduced system
involving modes $i$ and $j$. The corresponding characteristic
function is given by
$\Phi^{(k)}(\mu_{i},\mu_{j})=N_{k}(\alpha)\int d^{2}\mu_{k}
\langle\alpha|\hat{D}_{k}^{\dagger}(\mu_{k})|\alpha\rangle\Phi(\mu)$
with $N_{k}(\alpha)=\pi^{-1}(Q_{k}+1/2)\exp[(Q_{k}+1/2)^{-1}\left|
\alpha\right| ^{2}]$. Using Eq.~(\ref{eq:realvar1}), we obtain
$\Phi^{(k)}(\vec{\mu}
_{i},\vec{\mu}_{j})=\exp[-(\vec{\mu}_{i},\vec{\mu}_{j})V^{(k)}(\vec{\mu}
_{i},\vec{\mu}_{j})^{T}+D^{(k)}(\vec{\mu}_{i},\vec{\mu}_{j})^{T}]$,
where the CM $V^{(k)}$ is given by (\ref{eq:CMVk}) except for the
replacements
\begin{equation}
Q_{i(j)}\rightarrow
Q_{i(j)}-\frac{T_{j(i)}^{2}}{Q_{k}+1/2},\;\;\;T_{k} \rightarrow
T_{k}+\frac{T_{i}T_{j}}{Q_{k}+1/2}\label{eq:Qetero}
\end{equation}
and the drift $D^{(k)}$ is given by
\begin{align}
&  D^{(k)}=\frac{2i}{Q_{k}+1/2}\\
&  \times\left(  \alpha^{(R)}T_{j},-\alpha^{(I)}T_{j},(-)^{k+1}\alpha
^{(R)}T_{i},-\alpha^{(I)}T_{i}\right)  \nonumber
\end{align}
[Notice that $D^{(k)}$ depends on $\alpha$ and on the $i-j$ ordering too]. It
is evident that the distilled channel is still a Gaussian state with CM in the
standard form but with a non-zero drift. It follows that Alice ($i$) and Bob
($j$) can arrange the teleportation protocol of Section~\ref{teoria} with a
suitable additional displacement $\delta_{\pm}^{(k)}$ and with fidelity
$F_{\pm}^{(k)}$ given by (\ref{eq:FIDstform}) for a coherent input. For $k=1$
we have again $F_{\pm}^{(1)}\leq1/2$, while for $k=0,2$ we have an EPR channel
with a modified fidelity $F_{\pm}^{(k)}=[1+Q_{i}+Q_{j}\pm2T_{k}-(T_{i}\mp
T_{j})^{2}(Q_{k}+1/2)^{-1}]^{-1}$. In such a case ($k=0,2$) we have
$\delta_{\pm}^{(k)}=\alpha(T_{i}\mp T_{j})(Q_{k}+1/2)^{-1}$. In
Fig.~\ref{fig1} we report $F_{\pm}^{(2)}$ versus $t^{\prime}$ and for $\bar
{n}=0,1,10^{7}$ ($r=1+2.5\times10^{-7}$). When $\bar{n}=0$ we have
$F_{+}^{(2)}>1/2$ for $\pi<t^{\prime}<2\pi$ and $F_{-}^{(2)}>1/2$ for
$0<t^{\prime}<\pi$. However, when $\bar{n}\neq0$, the quantum character
survives only in small intervals before $t^{\prime}=2\pi$ and after
$t^{\prime}=0$ (or equivalently after $t^{\prime}=2\pi$). These fidelities not
only have a temperature-independent maximum value $[F_{\pm}^{(2)}]_{\max
}\simeq0.85$ greater than that of the traced out mode case, but also the
``quantum time intervals'' are larger and more robust with respect to
temperature. Moreover, as in the traced out mode case, the distilled channel
shared by the Stokes mode (1) and the mirror mode (0) has both types of EPR
correlations and therefore one can perform a further quantum teleportation
just after $t^{\prime}=2\pi$ ($\Longleftrightarrow$ $t^{\prime}=0$)\ in
addition to the one suggested in \cite{PRA}. Finally, in Fig.~\ref{fig2}, we
compare $F_{-}^{(0)}$ versus $t^{\prime}$ when the vibrational mode is
detected ($F_{-}^{(0,het)}$) with the corresponding fidelity when the same
mode is traced out ($F_{-}^{(0,tr)}$), for $\bar{n}=0,10^{5}$. The improvement
of the fidelity brought by the additional heterodyne measurement
($F_{-}^{(0,het)}\sim1$ for $\bar{n}=0$, except very close to $2m\pi$) is
impressive, meaning that heterodyning the mirror mode (and communicating the
result) allows to distill more EPR correlations between the two optical
sidebands. In other words, the heterodyne measurement allows to improve the
efficiency of the present scheme as an alternative source of two-mode
squeezing \cite{JOPB}.

\section{Conclusions}

\label{conclu}

In conclusion, we have presented an optomechanical system as a paradigm of
three-mode teleportation network. We have provided a thorough study of
entanglement and teleportation capabilities of this quantum channel. The
teleportation fidelities result improved by using heterodyne measurement at
the remaining mode and using the acquired information to distill a finer channel.

On one hand, our results could be useful to extend quantum information
processing towards macroscopic domain, by using e.g.
micro-opto-mechanical-systems \cite{stowe}. On the other hand, they could be
applied as well to all optical systems described by the Hamiltonian of
Eq.~(\ref{eq:Heff}) \cite{allopt}.

\begin{thebibliography}{99}
\bibitem{man}M. A. Nielsen and I. L. Chuang, \textit{Quantum Computation and
Quantum Information}, Cambridge University Press, Cambridge, 2000).

\bibitem {vai}L. Vaidman, Phys. Rev. A \textbf{49}, 1473 (1994); S. L.
Braunstein and H. J. Kimble, Phys. Rev. Lett. \textbf{80}, 869 (1998).

\bibitem {Tclon2}P. van Loock and S. L. Braunstein, Phys. Rev. Lett.
\textbf{87}, 247901 (2001).

\bibitem {telenet}P. van Loock and S. L. Braunstein, Phys. Rev. Lett.
\textbf{84}, 3482 (2000).

\bibitem {PRA}S. Mancini, \textit{et al.}, Phys. Rev. Lett. \textbf{90},
137901 (2003); S. Pirandola, \textit{et al.}, arXiv:quant-ph/0309078 (in press
on Phys. Rev. A).

\bibitem {PIN99}M. Pinard, \textit{et al.}, Eur. Phys. J. D \textbf{7}, 107 (1999).

\bibitem {JOPB}S. Pirandola \textit{et al.}, J. Opt. B: Quantum Semiclass.
Opt. \textbf{5}, S523-S529 (2003).

\bibitem {TIT99}I. Tittonen, \textit{et al.}, Phys. Rev. A \textbf{59}, 1038 (1999).

\bibitem {QO94}D. F. Walls and G. J. Milburn, \textit{Quantum Optics},
(Springer, Berlin, 1994).

\bibitem {entcirac}G. Giedke, \textit{et al.}, Phys. Rev. A \textbf{64},
052303 (2001).

\bibitem {CHI02}A. V. Chizhov, \textit{et al.}, Phys. Rev. A \textbf{65},
022310 (2002); J. Fiur\'{a}\v{s}ek, Phys. Rev. A \textbf{66}, 012304 (2002).

\bibitem {BRA99}S. L. Braunstein, \textit{et al.}, Phys. Rev. A \textbf{64},
022321 (2001).

\bibitem {PRLdu}L.-M. Duan, \textit{et al.}, Phys. Rev. Lett. \textbf{84},
2722 (2000).

\bibitem {stowe}T. D. Stowe, \textit{et al.}, Appl. Phys. Lett. \textbf{71},
288 (1997).

\bibitem {allopt}M. E. Smithers, and E. Y. C. Lu, Phys. Rev. A \textbf{10},
1874 (1974); A. Ferraro \textit{et al.}, e-print quant-ph/0306109.
\end{thebibliography}
\end{document}